\documentclass{article}
\usepackage{fix-cm}
\usepackage[T1]{fontenc}
\usepackage{finesid_arxiv,times}
\usepackage{amsmath,amssymb,booktabs,tabularx,microtype,xcolor,etoolbox}
\usepackage{graphicx,tikz}
\usetikzlibrary{arrows.meta,positioning}
\usepackage[colorlinks=true,allcolors=blue!55!black]{hyperref}
\newcommand{\method}{FineSID}

\newcommand{\E}{\mathbb{E}}
\title{Progressive Learning from Content\\to Relevance to Personalization\\for CTR Prediction}
\author{Zhiguang Yang \quad Yulin Yang \quad Fuyu Lv \quad Dan Ou\\\textnormal{Alibaba Group}\\\textnormal{\href{mailto:zgyang1996@gmail.com}{\texttt{zgyang1996@gmail.com}}}}
\begin{document} 
\maketitle
\raggedbottom

\begin{abstract}
Multimodal click-through rate (CTR) prediction is central to search systems, where ranking depends on item content, query relevance, and user preference. 
Existing approaches incorporate content through dense embeddings or coarse discrete IDs, yet effectively integrating fine-grained heterogeneous content with ID-based predictors remains challenging. 
We propose \method{}, a progressive learning framework that unifies content tokenization, query--item semantic pretraining, and personalized post-training via a shared fine-grained discrete representation. 
Each stage optimizes an explicit objective while preserving token-level access to content.
First, a pretrained multi-scale visual tokenizer and text tokenizers convert heterogeneous content into fine-grained Semantic IDs (SIDs), explicitly aligning them with the ID features. 
Second, full-parameter semantic pretraining under query--item relevance supervision establishes a robust matching prior, jointly optimizing the interactions of multimodal SIDs and conventional fields within a shared predictor.
Third, this initialized predictor is directly fine-tuned with user click supervision to capture implicit personalized preferences. 
Extensive offline and online evaluations demonstrate substantial gains in both standard CTR prediction and core business metrics. 
Furthermore, \method{} improves within-product creative discrimination, demonstrating its semantic generalization beyond conventional ID memorization.
\end{abstract}

\section{Introduction}

In billion-scale industrial search platforms like Taobao, accurate multimodal click-through rate (CTR) prediction is critical for both user experience and revenue.
The system must simultaneously model three progressive aspects: what an item contains (\emph{Content}), how it matches the query (\emph{Relevance}), and whether it appeals to the user (\emph{Personalization}). 
While industrial CTR predictors heavily rely on sparse identity (ID) features \citep{gui2023hiformer,zhu2025rankmixer,jiang2026tokenmixerlarge}, pure ID-based models struggle with cold-start items and fail to directly leverage the multimodal semantic knowledge available from modern Vision-Language Models (VLMs).

Integrating multimodal content into ID-based models remains a persistent challenge. 
One-stage joint training and two-stage continuous embedding fusion suffer from computation overhead and face distribution disparities between dense continuous features and the sparse ID space. 
Recently, Semantic IDs (SIDs) \citep{yu2022visualid,singh2024semantic} have been proposed to bridge this gap by quantizing global content embeddings into discrete codes. 
However, standard quantization schemes (e.g., Residual-Quantized Variational AutoEncoders or RQ-VAE) compress information aggressively, discarding spatial structures and textual details necessary for fine-grained semantic matching.

To address this, we propose \method{}, a progressive learning framework that exposes fine-grained multimodal content to a shared ID-based CTR predictor. 
\method{} structures learning across an explicit \emph{Item $\rightarrow$ Item--Query $\rightarrow$ Item--Query--User} progression:

\begin{itemize}
    \item \textbf{Stage I: Fine-grained Content Tokenization (\emph{Item}).} 
    Instead of relying on global compression, we preserve structural and semantic details by maintaining independent, field-reserved representations for different modalities. 
    We employ a pretrained multi-scale visual tokenizer \citep{tian2024var} alongside source-specific text tokenizers. 
    These extract multi-scale SIDs that map visual and textual details into the categorical ID space, ensuring the predictor retains token-level access to multimodal content.
    
    \item \textbf{Stage II: Semantic Pretraining (\emph{Item--Query}).}
    The shared predictor is trained under query--item relevance supervision. 
    Visual SIDs, textual SIDs, and conventional fields interact in a common prediction space. 
    By retaining fine-grained tokens rather than a single global embedding, the interaction network is guided to learn explicit cross-modal semantic alignment rather than simply memorizing global representations.
    
    \item \textbf{Stage III: Personalized Post-training (\emph{Item--Query--User}).} 
    The initialized shared predictor is then fine-tuned end-to-end with user click supervision. 
    This stage adapts the pretrained semantic interactions to incorporate personalized user behaviors and context.
\end{itemize}

We evaluate \method{} on the industrial Taobao platform and the public MMCTR dataset. 
Offline results on Taobao show that \method{} yields consistent AUC and GAUC improvements over the evaluated baselines, with notable gains on cold-start items. 
Furthermore, we evaluate within-product creative discrimination using the NSCTR metric \citep{yang2024pericr}. 
By comparing creatives sharing the exact same product ID, the results indicate that \method{} captures fine-grained content differences independent of conventional ID memorization. 
A completed online A/B experiment also reports positive changes in both user clicks and Gross Merchandise Value (GMV).

In summary, our key contributions are:
\begin{enumerate}
    \item We propose \method{}, a three-stage progressive learning framework that integrates multimodal content into ID-based CTR prediction via a discrete interface.
    \item We utilize multi-scale, fine-grained Semantic IDs to represent multimodal content. This approach aims to minimize information loss through reconstruction-oriented visual tokenization and source-specific text tokenization, helping the predictor capture spatial and textual details for explicit semantic alignment.
    \item We report offline comparisons, ablations, and online A/B outcomes. The results show consistent improvements in standard product-level CTR prediction, cold-start scenarios, and within-product creative discrimination.
\end{enumerate}

\section{Problem Formulation}

Let an exposure be defined as $x=(u,q,i,c)$, where $u$ denotes user and context features, $q$ a search query, $i$ an item ID, and $c$ its fine-grained multimodal content (e.g., displayed image and textual fields). 
Notably, different content (creatives) may belong to the exact same item.

\paragraph{Progressive Prediction Tasks.}
Our ultimate objective is to predict the user click probability:
\begin{equation}
 p_{\theta,\psi_{\mathrm{ctr}}}(x)=\Pr(y_{\mathrm{ctr}}=1\mid u,q,i,c),\qquad y_{\mathrm{ctr}}\in\{0,1\}.
\end{equation}
$\theta$ denotes shared feature-interaction parameters, and $\psi_{\mathrm{ctr}}$ is the task-specific CTR output head.

Instead of training from scratch, \method{} progressively learns $\theta$. We first pretrain the shared parameters $\theta$ under query--item relevance supervision $y_{\mathrm{rel}}$ :
\begin{equation}
 \pi_{\theta,\psi_{\mathrm{rel}}}(x)= \Pr(y_{\mathrm{rel}} \mid u,q,i,c),
\end{equation}
where $\psi_{\mathrm{rel}}$ is a dedicated relevance head. After establishing this semantic prior, the input interface and the shared network $\theta$ are entirely retained, and the model is fine-tuned on the personalized CTR task via click supervision.

\paragraph{Evaluation Scenarios.}
We evaluate the performance of the predictor under two distinct scenarios:
\begin{itemize}
    \item \textbf{Product-level CTR prediction:} Candidate exposures span distinct item IDs, testing overall ranking capacity.
    \item \textbf{Within-product creative discrimination:} The item ID $i$ is fixed, and candidates vary only in their displayed content $c$.
\end{itemize}

\section{FineSID}

\begin{figure}[t]
\centering
\includegraphics[width=\linewidth]{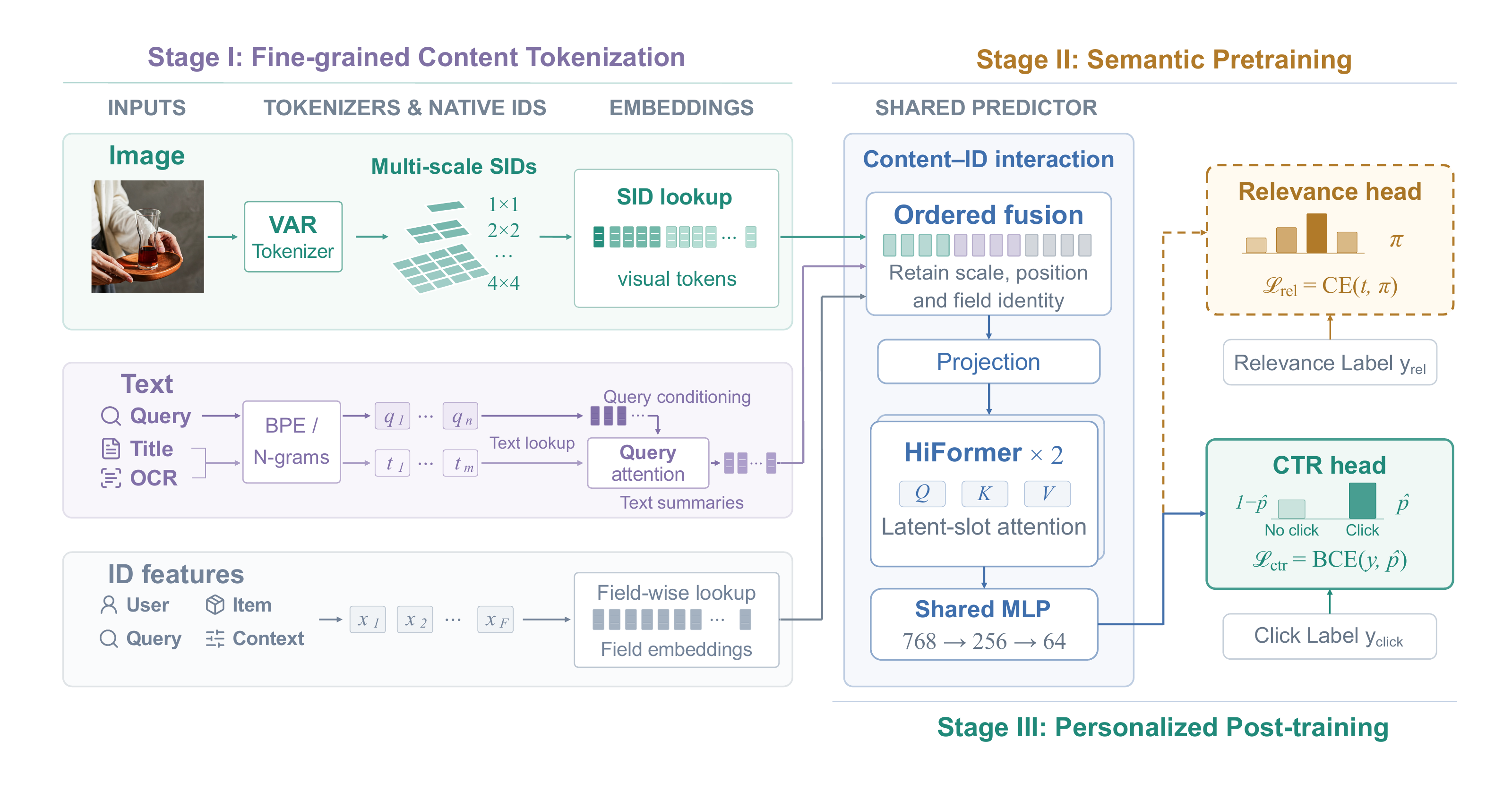}
\caption{
The overall architecture of the \method{} framework. 
The progressive learning process is decoupled into three stages. 
\textbf{Stage I: Fine-grained Content Tokenization} extracts fine-grained discrete representations, where images are converted into multi-scale visual tokens via a VAR tokenizer, and text fields are tokenized into discrete text IDs and mapped to embeddings through lookup tables. 
\textbf{Stage II: Semantic Pretraining} trains the entire shared predictor (including ordered fusion, latent-slot attention via HiFormer, and the shared MLP) using query--item relevance labels to establish a strong semantic matching prior. 
\textbf{Stage III: Personalized Post-training} fine-tunes the relevance-initialized shared predictor under user click supervision for final CTR prediction.
For simplicity, $y_{\mathrm{rel}}$ denotes the teacher-provided soft relevance target $t(x)$, $y_{\mathrm{click}}=y$, and $\pi$ and $\hat p$ denote the model's relevance distribution and click probability, respectively.
}
\label{fig:architecture}
\end{figure}

As illustrated in Figure \ref{fig:architecture}, \method{} is a progressive learning framework designed to integrate multimodal content into ID-based CTR prediction. 
The overall pipeline is decoupled into three progressive stages: (1) fine-grained content tokenization (Stage I), (2) full-parameter semantic pretraining (Stage II), and (3) personalized CTR post-training (Stage III). 
At the core of Stage II and Stage III lies a shared feature interaction architecture ($\theta$) that captures complex cross-modal relations.

\subsection{Stage I: Fine-Grained Content Tokenization}

In ID-based CTR predictors, continuous multimodal embeddings suffer from distributional misalignment and are often relegated to late fusion or auxiliary similarities. 
To seamlessly align multimodal content with the sparse ID space, \method{} discretizes it into Semantic IDs (SIDs), sharing the exact same lookup interface as categorical IDs for native feature interactions. 
Furthermore, our tokenization aims to limit compression-induced information loss by retaining the multi-scale spatial and structural details necessary for fine-grained ranking.

\paragraph{Multi-scale Visual Tokenization.} 
Specifically, we employ the pretrained multi-scale visual tokenizer of VAR \citep{tian2024var} to encode images into discrete visual SIDs. 
Its reconstruction-based training encourages high-fidelity preservation of visual content, supplying a spatially aware discrete interface before our downstream relevance and click objectives are applied. For an observed image $I$, we retain four native code maps across progressively finer scales:
\begin{equation}
 S^v_\ell\in\{0,\ldots,V-1\}^{h_\ell\times w_\ell},\qquad
 (h_\ell,w_\ell)\in\{(1,1),(2,2),(3,3),(4,4)\},
 \label{eq:pyramid}
\end{equation}
where $V$ is the visual codebook size. This yields 30 spatially indexed SIDs ($1+4+9+16$) that capture coarse-to-fine visual semantics. 
Because these discrete codes are extracted offline, deploying them provides deep semantic evidence without incurring the heavy latency of running an autoregressive image generator during click prediction.

During both Stage II and Stage III, the shared predictor maps each of these fixed visual SIDs to a task-trainable embedding:
\begin{equation}
 e^v_{\ell,p}=m_{\ell,p}E_v[S^v_{\ell,p}],\qquad
 z_v=[\operatorname{vec}(e^v_1);\operatorname{vec}(e^v_2);\operatorname{vec}(e^v_3);\operatorname{vec}(e^v_4)],
 \label{eq:visual}
\end{equation}
where $p$ indexes the spatial position within scale $\ell$, $m_{\ell,p}$ acts as a validity mask for missing codes, and $E_v$ is the randomly initialized, task-specific lookup table in our shared predictor (distinct from the tokenizer's original reconstruction codebook). 
The $\operatorname{vec}(\cdot)$ operation flattens the embeddings in a fixed spatial order, preserving both the scale and spatial position slots before they enter the feature interaction layers. 
Code zero is a valid ID and does not itself indicate padding.

\paragraph{Textual Tokenization.}
To complement the visual SIDs, we process textual content (e.g., search query, item title, and OCR texts) using Byte Pair Encoding (BPE) and n-gram tokenizers. 
Instead of compressing heterogeneous text into a single summary vector, we maintain independent, \emph{field-reserved} token sequences. Each sequence preserves its source identity and term position, exposing textual details through discrete lookups $E_t$, just as visual SIDs expose spatial image content. By maintaining separate vocabularies and fine-grained discrete interfaces for both modalities, Stage I ensures the shared predictor has comprehensive, token-level access to the multimodal content.

\subsection{Shared Feature Interaction Architecture ($\theta$)}

To effectively map the fine-grained multimodal SIDs and conventional ID features into a common semantic space, we deploy a shared feature interaction network (parameterized by $\theta$). This architecture is utilized in both Stage II and Stage III.

\paragraph{Ordered Fusion.} 
For textual content, we apply cross-attention with search-query embeddings as queries (Q) and each text field's token embeddings as keys and values (K, V), producing query-conditioned embeddings $a_t$. These attention outputs are passed to the shared predictor through ordered fusion.
We concatenate the conventional base-field embeddings $v_f$, the multi-scale visual representation $z_v$, and the query-conditioned text embeddings in a fixed order:
\begin{equation}
 z = [v_1;\ldots;v_F; z_v; \{a_t\}_t].
 \label{eq:fusion}
\end{equation}
This ordering preserves the scale and spatial slots of the visual codes and the field identities of the query-conditioned text representations before subsequent feature interactions.

\paragraph{Interaction Backbone.} 
The high-dimensional concatenated vector $z$ is first projected into a unified 768-dimensional global state. Since standard self-attention over thousands of raw tokens and ID fields is computationally prohibitive, we deploy a highly efficient latent-slot attention backbone inspired by HiFormer \citep{gui2023hiformer}. This network maps the global state into compact latent slots to capture high-order cross-feature interactions without the quadratic bottleneck. 

Finally, a Shared MLP configured with dimensions $768\rightarrow256\rightarrow64$ processes the refined global state to output the ultimate shared representation $h$. This compact representation $h$ is then fed into task-specific heads ($\psi_{\mathrm{rel}}$ or $\psi_{\mathrm{ctr}}$) in the subsequent progressive learning stages.

\subsection{Stage II: Full-Parameter Semantic Pretraining}
\label{sec:pretraining}

\paragraph{Memorization vs. Generalization.}
User satisfaction in search relies on two distinct dimensions: explicit \emph{query--item semantic relevance} and implicit \emph{personal preference} (e.g., visual appeal or price sensitivity). 
Conventional ID-based CTR models heavily rely on memorizing historical co-occurrences to capture the latter, but struggle with the former due to ID sparsity. 
Stage II injects a generalized semantic prior into the model. 
Crucially, without fine-grained content representation, models tend to devolve into merely memorizing global content embeddings. 
By preserving token-level details via \method{}, we guide the shared predictor to learn explicit cross-modal semantic alignment rather than resorting to rote memorization.

\paragraph{Relevance Objective.}
Using the complete exposure input $x$, we train the shared feature-processing network to judge whether the fine-grained candidate content satisfies the query intent.
For each eligible exposure $x$, a Qwen3.5 teacher model, fine-tuned on million-scale expert annotations, provides a soft relevance distribution $t(x)\in\Delta^3$ over four matching levels.
We optimize both the shared predictor parameters $\theta$ and a dedicated relevance head $\psi_{\mathrm{rel}}$ using a soft-label cross-entropy loss:
\begin{equation}
 \mathcal L_{\mathrm{rel}}=-\E_{x\sim\mathcal D_{\mathrm{rel}}}
    \sum_{k=1}^{4} t_k(x)\log \pi_{\theta,\psi_{\mathrm{rel}},k}(x).
\end{equation}
Importantly, relevance gradients update the \emph{entire} shared trainable parameter group $\theta$---including SID embeddings, categorical field lookups, interaction blocks, and normalization parameters. The associated normalization statistics are updated during training. 
Thus, the complete CTR backbone acquires a robust semantic prior. 

\subsection{Stage III: Personalized CTR Post-training}

Having established a generalizable semantic prior, the model must now learn the implicit, hard-to-articulate user preferences from behavioral logs. 
The primary advantage of \method{} is its seamless compatibility with existing industrial pipelines. 
Because multimodal content has already been aligned into the discrete ID space (Stage I) and the feature interactions have been semantically initialized (Stage II), Stage III requires no complex modality-fusion heuristics or contrastive losses. 

\paragraph{Click Objective.}
We initialize the CTR model directly with the pretrained shared parameters $\theta$ and its normalization states. 
We discard the relevance head, attach a randomly initialized CTR output layer $\psi_{\mathrm{ctr}}$, and jointly fine-tune the entire network with personalized user behavior:
\begin{equation}
 \mathcal L_{\mathrm{ctr}}=-\E_{(x,y)\sim\mathcal D_{\mathrm{ctr}}}
 \left[y\log p_{\theta,\psi_{\mathrm{ctr}}}(x)+(1-y)\log(1-p_{\theta,\psi_{\mathrm{ctr}}}(x))\right],
\end{equation}
where $y \in \{0, 1\}$ is the observed user click. 
Unlike standard embedding-only transfer methods that discard the pretrained dense network, \method{} transfers and fine-tunes all shared predictor parameters (full-parameter post-training). 
This simple, standard binary cross-entropy optimization proves sufficient to adapt the robust semantic interactions to highly personalized CTR predictions.

\section{Experiments}
\label{sec:experiments}

We conduct extensive experiments to evaluate the effectiveness of \method{}. We examine whether the proposed three-stage progressive learning framework can simultaneously improve standard product-level CTR prediction and fine-grained within-product creative discrimination.

\subsection{Datasets}
\label{sec:datasets}

\paragraph{Industrial Taobao Search.}
We use a chronological split of raw search impression logs: August 25--31, 2026 for CTR training and September 1, 2026 for offline evaluation. 
Each row is a candidate-item exposure with a binary click label and user, normalized-query, and item identifiers. 
We retain exposure rows rather than deduplicating repeated user--query--item interactions. 
Table~\ref{tab:dataset_statistics} summarizes the raw CTR populations; its training counts aggregate all seven days before distinct entities are counted. 
We count distinct nonempty user IDs, normalized query IDs, and item IDs over each complete split rather than summing daily distinct counts.
These statistics characterize the raw CTR data, before any method-specific feature or teacher-label availability conditions; they do not include the separate relevance-supervision corpus.

\begin{table}[!ht]
\centering\footnotesize
\setlength{\tabcolsep}{4pt}
\caption{Approximate size of the raw Taobao CTR splits in 2026. Exposure totals are rounded from row counts; User, Query, and Item are estimated distinct nonempty identifiers over the entire split.}
\label{tab:dataset_statistics}
\begin{tabular}{@{}llrrrr@{}}\toprule
Split & Dates & Exposures & Users & Queries & Items\\\midrule
Train & Aug 25--31 & 51.2B & 251M & 200M & 418M\\
Eval & Sep 1 & 453M & 30.2M & 15.4M & 1.40M\\\bottomrule
\end{tabular}
\end{table}

\subsection{Evaluation Metrics}
For product-level prediction, we report Area Under the Curve (AUC) and Group AUC (GAUC). GAUC is calculated by aggregating AUC computed within individual query groups, weighted by impression counts \citep{zhou2018din}: $\mathrm{GAUC}=\sum_g w_g\mathrm{AUC}_g/\sum_g w_g$. 
Furthermore, to evaluate the performance under data sparsity, we report metrics on a \emph{cold-start subset}, which contains items with fewer than 1,000 page views (PV) on the evaluation day (i.e., $\mathcal I_{\mathrm{cold}}=\{i:\mathrm{PV}(i)<1000\}$). 
For within-product creative ranking, we utilize the Normalized Simulated CTR (NSCTR) \citep{yang2024pericr}, detailed in Section~\ref{sec:creative_evaluation}. 
For the industrial Taobao dataset, results are expressed as improvements relative to the baseline (percentage points (pt) for AUC/GAUC, and relative percentage (\%) for NSCTR). Absolute AUC is reported for the public MMCTR dataset.

\subsection{Comparison Methods}
\label{sec:comparison_methods}

We compare \method{} with an ID-based baseline and several state-of-the-art approaches that integrate content semantics or relevance supervision into CTR prediction:
\begin{itemize}
    \item \textbf{HiFormer}~\citep{gui2023hiformer}: the foundational baseline using conventional ID features.
    \item \textbf{LEARN-style embedding}~\citep{jia2025learn}: directly augments the baseline with Qwen3.5 continuous content embeddings.
    \item \textbf{Visual ID}~\citep{yu2022visualid}: quantizes a visual embedding into discrete IDs.
    \item \textbf{RQ-VAE SID}~\citep{singh2024semantic}: uses residual-quantization IDs to compress content.
    \item \textbf{COINS}~\citep{zhao2026coins}: aligns content and collaborative information through semantic IDs for cold-start items.
    \item \textbf{PRECTR}~\citep{chen2025prectr}: couples personalized search relevance and CTR prediction.
    \item \textbf{Joint relevance + CTR}~\citep{chen2026joint}: jointly optimizes relevance and user engagement in a multitask ranking model.
\end{itemize}

\subsection{Taobao Offline Evaluation}

\begin{table}[!ht]
\centering\footnotesize
\setlength{\tabcolsep}{5pt}
\caption{Taobao offline improvements over the HiFormer plus ID features baseline. AUC and GAUC columns report absolute changes in percentage points (pt); NSCTR reports relative improvement (\%). `Cold` denotes items with evaluation-day PV $< 1000$.}
\label{tab:taobao_offline}
\begin{tabular}{@{}lrrrrr@{}}\toprule
Method & \shortstack{AUC\\(pt)} & \shortstack{GAUC\\(pt)} & \shortstack{Cold AUC\\(pt)} & \shortstack{Cold GAUC\\(pt)} & \shortstack{NSCTR\\(\%)}\\\midrule
HiFormer + ID features~\citep{gui2023hiformer} & -- & -- & -- & -- & --\\
LEARN-style embedding~\citep{jia2025learn} & +0.05 & +0.04 & +0.14 & +0.09 & +0.18\\
Visual ID~\citep{yu2022visualid} & +0.14 & +0.14 & +0.16 & +0.21 & +0.24\\
RQ-VAE SID~\citep{singh2024semantic} & +0.12 & +0.15 & +0.20 & +0.25 & +0.22\\
COINS~\citep{zhao2026coins} & +0.16 & +0.13 & +0.18 & +0.28 & +0.27\\
PRECTR~\citep{chen2025prectr} & +0.10 & +0.15 & +0.11 & +0.16 & --\\
Joint relevance + CTR~\citep{chen2026joint} & +0.04 & +0.03 & -0.04 & -0.07 & --\\
\method{} (Ours) & \textbf{+0.25} & \textbf{+0.26} & \textbf{+0.38} & \textbf{+0.34} & \textbf{+0.75}\\\bottomrule
\end{tabular}
\end{table}

As shown in Table~\ref{tab:taobao_offline}, \method{} achieves the largest gains across all reported metrics. Its overall AUC and GAUC improvements over the baseline are +0.25 and +0.26 points, respectively. Notably, these gains are amplified on the cold-start subset (+0.38 and +0.34 points), which is consistent with our hypothesis: when behavioral ID evidence is sparse, the predictor successfully leverages the robust semantic prior established by fine-grained content tokenization and relevance pretraining. Furthermore, the 0.75\% relative improvement in NSCTR demonstrates the framework's effectiveness in within-product creative discrimination, outperforming the evaluated global-quantization baselines.

\subsection{Ablation Study}

\begin{table}[!ht]
\centering\footnotesize
\setlength{\tabcolsep}{5pt}
\caption{Ablation study on Taobao. Each row reports its gain over the HiFormer baseline.}
\label{tab:taobao_ablation}
\begin{tabular}{@{}lrrrrr@{}}\toprule
Configuration & \shortstack{AUC\\(pt)} & \shortstack{GAUC\\(pt)} & \shortstack{Cold AUC\\(pt)} & \shortstack{Cold GAUC\\(pt)} & \shortstack{NSCTR\\(\%)}\\\midrule
\method{} (Full) & \textbf{+0.25} & \textbf{+0.26} & \textbf{+0.38} & \textbf{+0.34} & \textbf{+0.75}\\
w/o Relevance pretraining (Stage II) & +0.20 & +0.21 & +0.31 & +0.30 & +0.51\\
Pooling \method{} & +0.12 & +0.14 & +0.22 & +0.18 & +0.44\\
w/o Fine-grained SIDs (Stage I) & +0.03 & +0.02 & -0.03 & +0.04 & +0.02\\\bottomrule
\end{tabular}
\end{table}

To understand the contribution of each design component, we conduct ablations in Table~\ref{tab:taobao_ablation}. 
Removing the Stage II semantic pretraining reduces both overall and cold-start metrics, while driving the NSCTR improvement down from 0.75\% to 0.51\%. 
This highlights that end-to-end CTR training alone is prone to missing explicit query--item relevance signals. 
Furthermore, the \emph{Pooling \method{}} variant, which replaces the query-conditioned cross-attention and multi-scale slots with flat average pooling, yields smaller AUC gains over HiFormer: 0.12 percentage points versus 0.25 for the full model. 
This supports spatially indexed visual features and query-conditioned text embeddings over flat average pooling.
Crucially, entirely removing the fine-grained SIDs (reverting to pure ID features) substantially diminishes the improvements, resulting in a 0.03-percentage-point decrease in cold-start AUC relative to HiFormer and minimal NSCTR improvement (+0.02\%). 
This confirms that fine-grained content tokenization is a critical component for multimodal CTR improvements.

\subsection{Public Benchmark: MMCTR}

\begin{table}[!ht]
\centering\small
\caption{Performance on the public MMCTR dataset. Changes ($\Delta$) are absolute AUC differences from the HiFormer baseline, expressed in percentage points.}
\label{tab:mmctr}
\begin{tabular}{@{}lrr@{}}\toprule
Method & AUC & $\Delta$AUC (pt)\\\midrule
HiFormer + ID features & 0.9767 & Reference\\
LEARN-style embedding & 0.9823 & +0.56\\
COINS & 0.9825 & +0.58\\
\method{} (w/o relevance pretraining) & \textbf{0.9844} & \textbf{+0.77}\\\bottomrule
\end{tabular}
\end{table}

To validate the generalizability of our framework, we evaluate on the public MMCTR benchmark~\citep{zhu2021openbenchmarking}. Because the MMCTR dataset lacks the explicit query-item relevance labels required for Stage II, we specifically evaluate the \method{} variant \emph{without} relevance pretraining. As shown in Table~\ref{tab:mmctr}, even relying solely on Stage I (fine-grained content tokenization) and Stage III (CTR post-training), \method{} achieves the largest AUC gain (+0.77 pt) among all reported methods, further validating the utility of representing fine-grained SIDs as native categorical features.

\subsection{Within-Product Creative Evaluation}
\label{sec:creative_evaluation}

Standard product-level CTR evaluation often confounds a model's true semantic understanding with its capacity to memorize historical ID co-occurrences. To isolate the benefit of fine-grained multimodal representations, we evaluate \emph{within-product creative selection}. 
In this scenario, the item ID is identical across candidates; therefore, conventional ID-based memorization becomes uninformative. 
The predictor must therefore rely on resolving the fine-grained differences in the multimodal content tokens (e.g., varied images or titles for the exact same product) to accurately rank the creatives.

To quantify this, we utilize the Normalized Simulated CTR (NSCTR) introduced in Peri-CR~\citep{yang2024pericr}. For each logged exposure, the model scores candidate creatives of the same item and selects the highest-scoring one. A replay match occurs when that selection equals the creative actually displayed. Let $N_i$ be the original exposure count of item $i$, and let $M_i$ and $C_i$ be the matched exposure and click counts. For items where $M_i > 0$, the NSCTR is defined as:
\begin{equation}
 \mathrm{NSCTR}=\frac{\sum_i N_i(C_i/M_i)}{\sum_i N_i}.
 \label{eq:nsctr}
\end{equation}
This metric weights the matched CTR of each item by its original exposure count, effectively evaluating the quality of the selected creative. As reported in Table~\ref{tab:taobao_offline}, \method{} achieves a +0.75\% relative improvement in NSCTR, more than doubling the improvement of COINS (+0.27\%), demonstrating the framework's ability to generalize its semantic understanding of visual and textual details beyond simple ID memorization.

\subsection{Online A/B Testing}

\begin{table}[!ht]
\centering\small
\caption{Seven-day online A/B testing results on Taobao's main search traffic.}
\label{tab:online}
\begin{tabular}{@{}lr@{}}\toprule
Outcome & Relative change vs. Control\\\midrule
Item Page View (IPV / Clicks) & +0.21\%\\
Gross Merchandise Value (GMV) & +0.85\%\\\bottomrule
\end{tabular}
\end{table}

To verify the practical business value of the proposed framework, we deployed \method{} in the Taobao main search engine and conducted online A/B testing against the production baseline for seven days. As summarized in Table~\ref{tab:online}, \method{} yielded a +0.21\% improvement in Item Page Views (Clicks) and a +0.85\% increase in Gross Merchandise Value (GMV). Both improvements passed statistical significance testing. In a mature, billion-scale industrial platform like Taobao, such improvements represent meaningful business impact, demonstrating that the semantic prior learned through our progressive framework translates directly to user engagement and revenue gains.

\section{Related Work}

\paragraph{Multimodal Content Representations.}
Prior work incorporates visual content into CTR prediction and recommendation through visual encoders and task-aware representation learning \citep{ge2018image,liu2020category,chen2022debias,yang2024courier}.
Fine-grained vision--language modeling provides complementary approaches: FashionBERT jointly models image patches and text \citep{gao2020fashionbert}, FILIP uses token-level late interaction \citep{yao2022filip}, and MM-Rec models relations between image regions and text for news recommendation \citep{wu2022mmrec}.
These studies establish the value of content representations and fine-grained interactions across different tasks.
Industrial CTR deployment requires balancing content granularity with the computational cost of feature interaction. 
For \method{}, the relevant design question is how to retain separately accessible content details when constructing inputs to a highly efficient shared CTR predictor.

\paragraph{Semantic IDs as a Content Interface.}
Visual ID and residual-quantization semantic IDs encode content representations as discrete codes for downstream prediction \citep{yu2022visualid,singh2024semantic}.
COURIER also examines cluster IDs as a downstream interface for pretrained visual representations \citep{yang2024courier}.
More recent approaches explore content--collaborative alignment for cold-start prediction and prototype-composed semantic identifiers \citep{zhao2026coins,liu2026palette}.
These approaches motivate the use of discrete content interfaces, while the organization and granularity of the codes remain important design choices.
To retain explicit spatial structure, \method{} uses spatially indexed, multi-scale codes from VAR's reconstruction-trained tokenizer \citep{tian2024var}, together with field-specific textual tokens.
Its emphasis is on retaining structured content units whose task-trainable embeddings can interact with conventional features during relevance and CTR learning.

\paragraph{Relevance Supervision and CTR Transfer.}
Prior work combines relevance and behavioral supervision through representation pretraining, coupled prediction modules, and joint objectives.
SEMINAR learns reusable multimodal representations with multiple pretraining objectives, including query--item relevance \citep{shen2024seminar}.
PRECTR pretrains a relevance module before joint training with a CTR base model \citep{chen2025prectr}; PRECTR-V2 further develops a lightweight text encoder using relevance supervision and LLM distillation \citep{cao2026prectrv2}.
Joint relevance--engagement ranking instead optimizes the two tasks together and combines their outputs for ranking \citep{chen2026joint}.
\method{} emphasizes the scope of relevance pretraining and transfer: relevance supervision trains the shared SID embeddings, content-interaction layers, and prediction backbone, which are then transferred and fine-tuned for personalized CTR prediction.
This connects a fine-grained multimodal input interface with relevance-to-CTR transfer of the shared predictor.

\section{Discussion and Limitations}

While \method{} demonstrates significant improvements in multimodal CTR prediction and creative discrimination, it presents several limitations that suggest valuable directions for future research:

\paragraph{Towards End-to-End Semantic Tokenization.} 
Currently, the multi-scale visual tokenizer (Stage I) is pretrained for image reconstruction and kept frozen during relevance pretraining (Stage II). While this discrete interface effectively bridges modalities, it is not strictly optimized for search semantics. Future work could explore jointly fine-tuning the visual tokenizer alongside the relevance pretraining task, enabling an end-to-end semantic tokenization process that could yield a higher upper bound for cross-modal matching.

\paragraph{Broader Generalization and LLM Integration.} 
We have validated the effectiveness of the pretrained semantic prior primarily through personalized CTR post-training (Stage III). However, whether this progressive pretraining paradigm can seamlessly generalize to other downstream search tasks (e.g., early-stage retrieval, conversion rate prediction) remains to be fully explored. Integrating the fine-grained SID representation with Large Language Models (LLMs) to build a more comprehensive, universal foundational model for search systems is an exciting open direction.

\paragraph{Computational Overhead in Ranking vs. Retrieval.} 
Replacing a single global image embedding with 30 multi-scale tokens inevitably increases the width of the concatenated feature vector and the computational cost of the initial bottom-layer projection. 
However, because the heavy Stage I token extraction is executed entirely offline, online inference only incurs the cost of extra discrete ID lookups and this widened initial projection, before the representation is mapped into fixed-capacity latent slots. 
In practice, this additional computational overhead is negligible; our online deployment confirmed no significant increase in serving latency during the late-stage ranking phase. 
Nonetheless, extending this fine-grained tokenization framework to extreme low-latency early-stage retrieval systems (which process millions of candidates) remains an area for future optimization.

\section{Conclusion}

In this paper, we introduced \method{}, a three-stage progressive learning framework that integrates multimodal content into ID-based CTR prediction. By discretizing multimodal content into fine-grained Semantic IDs, the framework seamlessly unifies content tokenization, semantic relevance pretraining, and personalized post-training within a shared ID-based architecture. This decoupled design enables the predictor to establish a robust relevance-informed semantic prior before adapting to implicit, personalized user behaviors. 

Offline evaluations on the industrial Taobao dataset and the public MMCTR benchmark, alongside completed online A/B testing, demonstrate the effectiveness of our approach. Specifically, \method{} not only improves standard product-level CTR prediction but also exhibits strong capability in within-product creative discrimination. These results validate that retaining fine-grained multi-scale structural details through reconstruction-oriented visual tokenization is crucial for unlocking the full potential of multimodal semantics in industrial search systems.

\section*{Reproducibility and AI Assistance}
To ensure reproducibility, detailed formulations of our evaluation metrics and comprehensive implementation details (including network configurations and hyperparameter settings) are provided in the Appendix. 

In accordance with the ICLR 2027 policy on AI assistance, we disclose that Large Language Models (LLMs) were utilized during the preparation of this manuscript. Specifically, AI tools were employed to: (1) aid in the retrieval and discovery of related literature; (2) assist with research ideation and execution; (3) draft initial sections of the paper; and (4) polish the final writing, organize references, and format equations. All AI-generated content was thoroughly reviewed and critically revised by the human authors. The experimental designs, data analyses, and ultimate scientific claims remain entirely the responsibility of the authors.

{\small\raggedright
\bibliography{references}
\bibliographystyle{iclr2027_conference}}

\appendix
\section{Measurement Details}
For a specific evaluation group $g$ (e.g., a query group) containing $n_g^+$ positive samples (clicks) and $n_g^-$ negative samples (non-clicks), let $W_g$ denote the number of correctly ordered positive--negative pairs plus half of the tied pairs. The exact AUC for this group is calculated as $\mathrm{AUC}_g = W_g / (n_g^+ n_g^-)$. For the set of valid groups $\mathcal G_+=\{g:n_g^+n_g^->0\}$, the Group AUC (GAUC) is defined as:
\begin{equation}
 \mathrm{GAUC}=\frac{\sum_{g\in\mathcal G_+}n_g\mathrm{AUC}_g}
                    {\sum_{g\in\mathcal G_+}n_g},\qquad n_g=n_g^++n_g^-.
\end{equation}
The denominator effectively weights the contribution of each valid group by its total impression count $n_g$. Note that while Item GAUC compares exposures across different users and contexts for a given item, it is fundamentally distinct from NSCTR (defined in Section~\ref{sec:creative_evaluation}), which specifically evaluates the fine-grained ranking of different creatives within the exact same item ID.

\section{Implementation and Reproducibility Details}
\paragraph{Tokenizer and Feature Configurations.} 
The framework explicitly distinguishes between the upstream image tokenizer and the downstream trainable SID lookup tables. For Stage I visual tokenization, we utilize a pretrained VAR checkpoint. We retain code maps corresponding to resolutions of $1\times1$, $2\times2$, $3\times3$, and $4\times4$, yielding exactly 30 spatially indexed discrete codes ($1+4+9+16$) per image. Note that this multi-scale retention is an explicit subset of the source tokenizer's full scale schedule. Missing visual codes are handled via a dedicated validity mask $m_{\ell,p}$. Concurrently, text sources (e.g., query, title) are encoded using distinct BPE and n-gram vocabularies to preserve source-specific sequence identities.

\paragraph{Network and Full-Parameter Transfer (Stage II to Stage III).} 
The shared CTR predictor incorporates 65 dynamic categorical embedding tables alongside the multimodal SIDs. During the Stage III personalized post-training, the model undergoes a strict full-parameter transfer protocol. Specifically, all 197 dense network states and normalization entries learned during the Stage II relevance pretraining are exactly retained and loaded. The sparse Adagrad accumulators are reset to avoid carrying relevance-stage adaptive learning-rate statistics into CTR training, and a fresh training step is initialized for the CTR task.

\paragraph{Training Hyperparameters and Hardware.}
The shared predictor's dense parameters and sparse embeddings are optimized using Adagrad with a fixed learning rate of $0.01$. Training is distributed across 64 NVIDIA H20 GPUs, with a per-GPU batch size of $4{,}096$ and a global batch size of $262{,}144$. No learning-rate warmup or decay schedule is used. For the multi-scale VAR tokenizer, the discrete visual codebook size is fixed at $V = 4{,}096$.

\end{document}